# Ultra-high Sensitivity Moment Magnetometry of Geological Samples Using Magnetic Microscopy

Eduardo A. Lima[1]*, Benjamin P. Weiss[1]

[1]Department of Earth, Atmospheric, and Planetary Sciences, Massachusetts Institute of Technology (MIT), 77 Massachusetts Avenue, Cambridge, MA 02139, USA

Corresponding author: Eduardo A. Lima, 77 Massachusetts Avenue, Room 54-724, Cambridge, MA 02139, USA, tel.: (617) 324-2829, fax: (617) 258-7401, email: limaea@mit.edu

**Key Points**

- We describe measurements of geological samples with magnetic moments up to 1000x weaker than the detection limit of standard superconducting rock magnetometers

- Detection and isolation of magnetic contamination in ultra-weak samples significantly improves reliability and accuracy compared to bulk moment measurements

- The accuracy is determined by the signal-to-noise ratio of the magnetic field maps and by the contribution from non-dipolar magnetization.

**Paleomagnetically useful information is expected to be recorded by samples with moments up to three orders of magnitude below the detection limit of standard superconducting rock magnetometers. Such samples are now detectable using recently developed magnetic microscopes, which map the magnetic fields above room-temperature samples with unprecedented spatial resolutions and field sensitivities. However, realizing this potential requires the development of techniques for retrieving sample moments from magnetic microscopy data. With this goal, we developed a technique for uniquely obtaining the net magnetic moment of geological samples from magnetic microscopy maps of unresolved or nearly unresolved magnetization. This technique is particularly powerful for analyzing small, weakly magnetized samples such as meteoritic chondrules and terrestrial silicate crystals like zircons. We validated this technique by applying it to field maps generated from synthetic sources and also to field maps measured using a superconducting quantum interference device (SQUID) microscope above geological samples with moments down to $10^{-15}$ Am$^2$. For the most magnetic rock samples, the net moments estimated from the SQUID microscope data are within error of independent moment measurements acquired using lower sensitivity standard rock magnetometers. In addition to its superior moment sensitivity, SQUID microscope net moment magnetometry also enables the identification and isolation of magnetic contamination and background sources, which is critical for improving accuracy in paleomagnetic studies of weakly magnetic rocks.**

1. Introduction

Until recently, the most sensitive magnetometers in the geosciences were capable of measuring the natural remanent magnetization (NRM) down to a limiting resolution of ~0.1-1×$10^{-12}$ Am$^2$. However, it has long been recognized that geological samples should be able to provide useful paleomagnetic records for moments at least several orders of magnitude below this threshold [*Kirschvink*, 1981]. Examples of such samples include chondrules and inclusions in chondritic meteorites [*Lappe et al.*, 2013; *Lappe et al.*, 2011; *Uehara and Nakamura*, 2006], which may contain records of magnetic fields in the solar nebula, and detrital zircon crystals, which might provide records of the earliest history of the Earth's magnetic field [*Tarduno et al.*, 2015; *Weiss et al.*, 2015].

Over the last two decades, a new technique for measuring the magnetic field above room temperature samples has been developed called superconducting quantum interference device (SQUID) microscopy [*Fong et al.*, 2005; *Weiss et al.*, 2007b]. By employing small (typically <100 μm) pickup loops brought extremely close to the samples, these instruments are capable of mapping the vertical component of the sample magnetic field with resolutions as low as 10 pT at spatial resolutions of 150 μm or better. Because the three components of the magnetic field measured in source-free space are interrelated by Gauss's Law and Ampère's Law, the two transverse field components can be uniquely calculated with high accuracy from these data, yielding the full vector field in a plane above the sample [*Lima and Weiss*, 2009].

The main application of this technique has been to infer the fine-scale magnetization distribution within geological samples [*Fu et al.*, 2012b; *Gattacceca et al.*, 2006; *Lima et al.*, 2013; *Oda et al.*, 2011; *Weiss et al.*, 2007a]. However, recovering magnetization distributions from field data is generally nonunique [*Baratchart et al.*, 2013; *Weiss et al.*, 2007b]. By comparison, magnetic field maps of a spatially unresolved (i.e., purely dipolar) sample can be used to uniquely retrieve the net magnetic moment of the source [*Lima et al.*, 2006; *Weiss et al.*, 2007b]. Our initial demonstrations of this approach have already established that moments as weak as $10^{-13}$ to $10^{-14}$ Am$^2$ can be retrieved, with the moments of relatively strongly magnetized samples obtained from SQUID microscopy and standard SQUID rock magnetometry in agreement [*Weiss et al.*, 2007b; *Weiss et al.*, 2008]. Furthermore, we recently successfully applied this technique as part of a comprehensive study of chondrules isolated from primitive chondritic meteorites that provided the first reliable paleointensities of fields in the solar nebula [*Fu et al.*, 2014], and to zircons from Bishop Tuff that provided accurate paleointensity measurements of the recent geomagnetic field [*Fu et al.*, submitted].

Here we present the first comprehensive demonstration of the accuracy and utility of this technique, its computational methodology and limitations, and its application to a typical paleomagnetic measurement suite involving progressive demagnetization. We begin in Section 2 by demonstrating that samples with moments at least 100 times weaker than those detectable with standard SQUID rock magnetometers contain paleomagnetically meaningful information. Having demonstrated why we are developing the moment magnetometry technique for ultrasensitive instruments like SQUID microcopes, we describe the technique in Section 3. Then, in Section 4, we validate it through its application to controlled magnetization distributions for synthetic and natural samples and comparison to measurements from standard rock magnetometers. We then apply the technique to the measurement of samples that are inaccessible to standard rock magnetometers, demonstrating the detection of moments as weak as $1\times10^{-15}$ Am$^2$.

## 2. The need for ultra-high sensitivity moment magnetometry

Standard SQUID rock magnetometers like the 2G Enterprises 755 Superconducting Rock Magnetometer (SRM) have moment sensitivities of $1\times10^{-12}$ Am$^2$ due to limitations in SQUID noise. In practice, background variations in the moment of sample holders will limit this to $1\times10^{-8}$ Am$^2$ or even higher unless special precautions are taken to use nonmagnetic materials [*Kirschvink et al.*, 2015]. Until recently, these limitations in instrument sensitivity have been the main factor determining the samples with the weakest magnetic moments that have been used for paleomagnetic measurements. For example, standard-sized (cm-scale) samples of lithologies with weak magnetizations ($\leq10^{-3}$ Am$^{-1}$) like carbonates and lunar basalts, as well as smaller (~0.1 mm) samples of geological materials like single silicate crystals [*Tarduno et al.*, 2015] and chondrules [*Fu et al.*, 2014] can have moments several orders of magnitude below $10^{-12}$ Am$^2$, particularly after laboratory demagnetization.

We next use a simple analysis to demonstrate that there should be paleomagnetically meaningful information carried by ferromagnetic grain assemblages with moments well below $10^{-12}$ Am$^2$. In particular, we estimate the smallest magnetic moment for a grain assemblage that would accurately constrain the paleointensity or the paleodirection of an ancient magnetizing field ***B***. The error in the recorded paleodirection is defined as the angle between the net moment (i.e., resultant) of the assemblage and the ancient field direction. The error in paleointensity is defined as the fractional deviation of the efficiency, *e* (i.e., ratio of the net moment to the saturation moment), relative to that typically observed for large numbers of grains. We consider two different paleofield strengths that produce *e* = 0.015 and 0.15, which empirically are observed for typical grain assemblages carrying a thermoremanence acquired in fields *B* of 50 and 500 µT, respectively [*Yu*, 2010].

We consider an assemblage of identical single domain magnetite crystals with uniaxial anisotropy and spontaneous magnetic moments $M_s$ with orientations distributed uniformly across the surface of the unit sphere. An extreme lower limit on the weakest most useful magnetic moment is set by the spontaneous moment of one spherical single domain grain with radius just above the superparamagnetic threshold (~25 nm) [*Butler and Banerjee*, 1975], for which $m \sim 3\times10^{-17}$ Am$^2$. Because this grain can be magnetized only in two directions and always has the same spontaneous moment intensity, it can record paleofield directions with errors up to 90° and essentially cannot record paleointensities.

To obtain a more meaningful lower limit, we use a Monte Carlo simulation to estimate the minimum number of grains, $n_{min}$, that must be measured to achieve on average an angular error of $\overline{\theta_e}$ = 10° in the paleofield direction $\hat{B}$ and an error in the paleofield intensity of $\overline{m_e}$ = 20%. We can see that for this uniform angular distribution of grain axes, $0 < e < 0.5$. We draw $n$ grains at random from this population and calculate their resultant, $\vec{R}$, for $n$ ranging from 1-10$^8$. We then compare the direction and magnitude of $\vec{R}(n)/n$ to that $\vec{R}(\infty)/\infty = e\hat{B}$ and compute the directional error:

$$\theta_e = \text{acos}[\hat{B} \cdot \vec{R}(n)/n]$$

and the paleointensity error:

$$m_e = 1 - \frac{\vec{R}(n)/n}{e}$$

We then repeated each of these experiments for 10,000 trials to calculate the mean values $\overline{\theta_e}$ and $\overline{m_e}$. This number of trial repetitions was found to insure convergence of the final estimated values of $\overline{\theta_e}$ and $\overline{m_e}$ to better than 1% of the true means.

For $e \sim 0.015$ and 0.15, we find that $n_{min}$ = ~70,000 and 800 for $\overline{\theta_e}$ = 10° and ~25,000 and ~250 for $\overline{m_e}$ = 20% (Fig. 1). Therefore, these intuitively specified directional and intensity error limits yield very similar values of $n_{min}$. Given the single grain moment $m$ above, these $n_{min}$ values correspond to net moments of ~2 and 0.2×10$^{-14}$ Am$^2$ for $\overline{\theta_e}$ = 10° and ~8 and 0.8×10$^{-15}$ Am$^2$ for $\overline{m_e}$ = 20%. An analytical study using Langevin theory for detrital remanent magnetization carried by single domain magnetite grains by Kirschvink [1981] estimated a minimum net moment of 6×10$^{-14}$ Am$^2$ for grains with radii about twice those considered here and for maximum mean directional errors of 5°. A recent analytical approach by Berndt et al. [2016] of magnetite grain assemblages found that paleointensity errors exceeded 20% and paleodirectional errors exceeded 20° for net moments of ~10$^{-15}$–10$^{-14}$ Am$^2$. In summary, three very different independent analyses by this study, Kirschvink [1981], and Berndt et al. [2016]

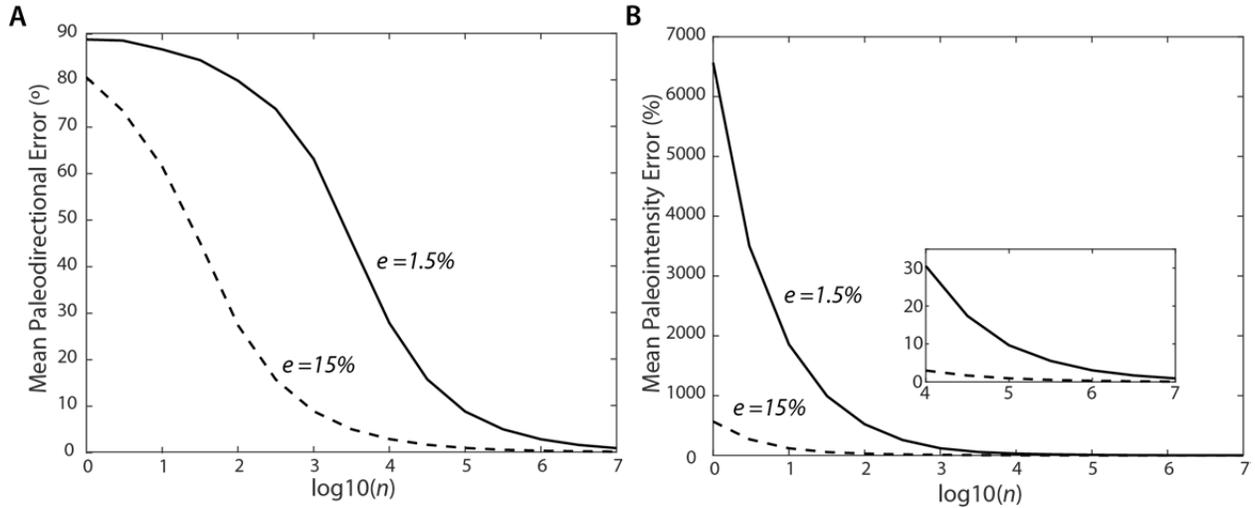

**Figure 1:** Paleodirectional and paleointensity errors associated with measuring the net moments of different numbers of single domain ferromagnetic grains with uniaxial easy axes distributed uniformly over the sphere. An underlying distribution of $10^8$ grains was magnetized to an efficiency $e$ of 1.5% or 15% in the paleofield direction (i.e., the net moment relative to the saturation remanence when all grains are magnetized in the same hemisphere centered around the paleofield direction $\hat{B}$). Then, $n$ grains were chosen at random and their net moments computed and compared to that expected for $n \to \infty$. **(A)** Mean paleodirectional errors $\overline{\theta_e}$. **(B)** Mean paleointensity errors $\overline{m_e}$. Inset shows mean paleointensity errors for $n \geq 10^4$.

have found that natural samples should preserve paleomagnetically useful information down to natural remanent moments of $10^{-15}$–$10^{-14}$ Am$^2$, 100 to 1000 times below that measurable with standard superconducting rock magnetometers.

## 3. Description of net moment technique

Having, demonstrated the need for high-sensitivity moment magnetometry, we now describe our technique in detail. We focus on the computation of net magnetic moments from measurements of the magnetic field of geological samples whose external field is close to that of a magnetic dipole. Dipole moments are powerful ways to represent experimental data because (I) they are the elemental building blocks of magnetization distributions; (II) magnetic fields of distant sources behave as those of single magnetic dipoles; (III) the external magnetic field produced by specific source geometries can be very close or even identical to that of a magnetic dipole [*Collinson*, 1983; *Jackson*, 1999; *Reitz et al.*, 2008]. We assume that the samples are analyzed in a zero-field environment — for instance, inside a magnetically shielded room or container — such that no observable induced magnetization component is present and background remanent magnetization is near zero due to the use of low-moment sample holders. These simplifications make the moment magnetometry problem for magnetic microcopy far more tractable than magnetic surveys of crustal magnetization (e.g., [*Parker*, 1991]).

Regarding (I), magnetization is a vector quantity that is defined as the macroscopic average of dipole moments in a small volume element. Thus, magnetic dipoles are fundamentally connected to magnetization distributions, which are ultimately composed of individual electron magnetic moments. As for (II), the magnetic field of any source of finite size approaches asymptotically the field of a magnetic dipole as the distance between the source and the observer (e.g., magnetic sensor) increases. Expressing the external magnetic field of such a source as an expansion of spherical harmonics field terms shows that at large distances the dipole term prevails over higher order multipoles [*Jackson*, 1999]. Owing to the orthogonality of the spherical harmonics basis, none of the higher order terms affects the dipole moment (i.e., they all have zero magnetic dipole moment). Interestingly, the coefficient of the dipole term in the spherical harmonics expansion corresponds to the integral of the magnetization (net magnetic moment) [*Jackson*, 1999; *Stratton*, 2007]. Moreover, the dipole moment term is invariant with respect to the origin of the spherical harmonics expansion [*Epton and Dembart*, 1995], which is particularly important for intricate magnetizations distributions for which the choice of a suitable origin for the expansion may not be obvious.

Finally, (III) stems from the fact that certain symmetrical source configurations yield fields external to the magnetization distributions that are identical to (e.g., uniformly magnetized sphere) or very close to (e.g., uniformly magnetized cylinder with specific aspect ratio) the field of a magnetic dipole even at somewhat small distances.

Next, we show how the magnetic moment can be recovered from magnetic field measurements for different experimental configurations.

## 3.1 - Recovering the moment

The field of a magnetic dipole is given by

$$\vec{B}(\vec{r}) = \frac{\mu_0}{4\pi} \left\{ \frac{3\vec{m} \cdot (\vec{r} - \vec{r}')}{|\vec{r} - \vec{r}'|^5} (\vec{r} - \vec{r}') - \frac{\vec{m}}{|\vec{r} - \vec{r}'|^3} \right\}, \qquad (1)$$

where $\vec{m}$ is the magnetic moment, $\vec{r}$ and $\vec{r}'$ represent the positions of the sensor and of the dipole, respectively, and $\vec{B}$ is the magnetic field. In the absence of noise, we only need one set of measurements of the three components of the magnetic field of a dipolar source at a known distance $[B_x(r), B_y(r), B_z(r)]$ to recover the net moment:

$$\vec{m} = \frac{4\pi}{\mu_0}|\vec{r}-\vec{r}'|^3\left\{\frac{3}{2}\frac{\vec{B}(\vec{r})\cdot(\vec{r}-\vec{r}')}{|\vec{r}-\vec{r}'|^2}(\vec{r}-\vec{r}')-\vec{B}(\vec{r})\right\}, \qquad (2)$$

given that for known $\vec{r}-\vec{r}'$ the moment $\vec{m}$ is completely specified by $\vec{B}$. [The expression above can be easily obtained from (1) the after some manipulation making use of vector identities.]

Alternatively, instead of using a single measurement of the vector magnetic field, we can also recover the moment from three measurements of a single component of the field taken at suitable positions (e.g., the *z*-component of the magnetic field measured at three points on a horizontal plane above the source):

$$\begin{bmatrix}B_z(x_1,y_1,h)\\B_z(x_2,y_2,h)\\B_z(x_3,y_3,h)\end{bmatrix} = \frac{\mu_0}{4\pi}\begin{bmatrix}\frac{3x_1 h}{(x_1^2+y_1^2+h^2)^{5/2}} & \frac{3y_1 h}{(x_1^2+y_1^2+h^2)^{5/2}} & \frac{2h^2-x_1^2-y_1^2}{(x_1^2+y_1^2+h^2)^{5/2}}\\ \frac{3x_2 h}{(x_2^2+y_2^2+h^2)^{5/2}} & \frac{3y_2 h}{(x_2^2+y_2^2+h^2)^{5/2}} & \frac{2h^2-x_2^2-y_2^2}{(x_2^2+y_2^2+h^2)^{5/2}}\\ \frac{3x_3 h}{(x_3^2+y_3^2+h^2)^{5/2}} & \frac{3y_3 h}{(x_3^2+y_3^2+h^2)^{5/2}} & \frac{2h^2-x_3^2-y_3^2}{(x_3^2+y_3^2+h^2)^{5/2}}\end{bmatrix}\begin{bmatrix}m_x\\m_y\\m_z\end{bmatrix}, \qquad (3)$$

where we assume that the dipole is located at the origin of the coordinate system (without loss of generality), and that the field is measured at three points on the plane $z = h$ parallel to the sample: $(x_1, y_1, h)$, $(x_2, y_2, h)$, and $(x_3, y_3, h)$. Such points must be chosen so as to yield a non-singular, invertible matrix when solving the system of linear equations (3). This can be easily accomplished by avoiding (i) points where $B_z = 0$; (ii) three points that are greatly clustered; (iii) three points that lie on a line (e.g., $x = y$, $x = 0$, $y = 0$) or on the circle of radius $h\sqrt{2}$ centered about the origin.

Under real experimental conditions, sensor noise will degrade the vector magnetic field measurement $\vec{B}$ in (2) or the three measurements of the *z*-component of $\vec{B}$ in (3), thereby affecting the recovered moment by adding a spurious component. To ameliorate this problem, an average magnetic moment estimate can be computed using (2) for repeated measurements of $\vec{B}$ at the same distance from the source. In the single-component approach, we can combine a larger number of measurements to obtain an overdetermined system of linear equations whose approximate solution can be found via the method of least squares.

Whereas these two approaches are very straightforward due to the fact they are linear in the recovered three components of the moment, the accuracy in recovering the moment is directly related to how well known is the position of the sensor relative to the source. For this reason, such methods tend to perform better when the sensor-to-sample distance is large compared to the sample size, such that small uncertainties in the relative position do not noticeably impact accuracy (similarly, sensors with large sensing areas or volumes tend to minimize the influence of position uncertainty, owing to averaging effects).

In scanning magnetic microscopy, the magnetic sensor is typically brought as close as possible to the sample in order to maximize sensitivity and achieve superior spatial resolution. Errors in assessing the relative distance between the measurement positions and the sample location may not be negligible, particularly because the exact location of the equivalent dipole in a given geological sample is usually not known. Therefore, a different approach is required to accurately recover the net moment.

In essence, we generalize the least-squares method for solving the linear system (3) to account for the uncertainty in the dipole location. Specifically, we assume a dipole that is no longer located at the origin but at the coordinate $\vec{r}' = (x_0, y_0, 0)$ (we take the *z*-coordinate to be zero without loss of generality). Measurements of the field component normal to the sample, $B_z$, are again taken at a plane parallel to the *x-y* plane and distance *h* above it. We then have to find six parameters in total, three of which (spatial coordinates of the dipole) exhibit a non-linear dependence with $B_z$. Thus, we have transformed our linear least-squares problem into a nonlinear one. Notice that $B_z$ still preserves its linear dependence on the three remaining parameters (i.e., the components of the dipole moment) such that the least-squares problem is actually of mixed form or separable. We also consider that the magnetic field measurements are taken at positions on an evenly spaced rectangular grid.

We draw attention to the fact that using a single component of the magnetic field to determine the magnetic moment does not lead to loss of information or to a decrease in accuracy.

Maxwell's equations establish that the magnetic field in a region devoid of sources (i.e. outside the sample, where we take measurements of the field) can be completely represented by the gradient of a scalar function satisfying Laplace's equation. This scalar potential means that all three field components are tightly interconnected and that a single component essentially carries all the information about the full vector field [*Lima and Weiss*, 2009]. Notice that these relationships hold for the magnetic field measured and computed *on surfaces* rather than in a pointwise manner. Despite a recent unsubstantiated assertion to the contrary [*Cottrell et al.*, 2016], this fact has been recognized and exploited in geophysics since as early as 1945 [*Vestine and Davids*, 1945], and is extensively used in magnetic surveys from the local to the planetary scale [*Blakely*, 1996; *Purucker*, 1990; 2008]

In mathematical terms, we wish to find the parameter vector $\boldsymbol{p} = (x_0, y_0, h, m_x, m_y, m_z)^T = (\boldsymbol{x}, \boldsymbol{m})^T$, that minimizes the objective (cost) function defined as the residual sum of squares between the experimental magnetic field data and the dipole model field computed at the same locations (the symbol $^T$ denotes the transpose of a vector or matrix). That is to say:

$$\underset{\boldsymbol{x}, \boldsymbol{m} \in \mathbb{R}^3}{\text{minimize}} \quad \|\boldsymbol{b}_z - \boldsymbol{G}(\boldsymbol{x})\boldsymbol{m}\|_2^2, \qquad (4)$$

where $\boldsymbol{G}$ stands for the geometry matrix, $\boldsymbol{b}_z$ is the vector with the measurements of the *z*-component of the magnetic field on the planar grid, and $\|\cdot\|_2^2$ denotes the Euclidean norm (2-norm) squared.

We can use two different methods to solve (4). Case 1: a nonlinear optimization algorithm searches a six-dimensional parameter space for the optimal parameter vector $\boldsymbol{p}$ that minimizes the residuals. Case 2: we take advantage of the linear relationship between the magnetic field and the magnetic moment (i.e., the separability of the nonlinear least-squares problem [*Golub and Pereyra*, 2003]) to split the solution into linear and nonlinear parts. Here, a nonlinear optimization algorithm searches instead a reduced three-dimensional space corresponding to the $\boldsymbol{x}$ parameter vector. For each iteration of the optimization algorithm, a linear least-squares problem is solved to find the moment vector $\boldsymbol{m}(\boldsymbol{x})$ for the particular geometry matrix $\boldsymbol{G}(\boldsymbol{x})$:

$$\underset{\substack{x \in \mathbb{R}^3 \\ \text{subject to } m(x)=G^+(x)b_z}}{\text{minimize}} \quad \|b_z - G(x)m(x)\|_2^2, \quad (5)$$

where $G^+(x) = \left(G(x)^T G(x)\right)^{-1} G(x)^T$ is the pseudoinverse of $G(x)$. (That is to say, for each set of values for $x$, $y$, and $h$, chosen by the optimization algorithm we solve for $[m_x, m_y, m_z]$.)

One possible interpretation of this problem is recognizing that the components of the magnetic moment are the coefficients in a linear combination of scalar-valued functions $G_1(x,y,h)$, $G_2(x,y,h)$, and $G_3(x,y,h)$ that best approximates the field data:

$$b_z(x,y,h) = m_x G_1(x-x_0, y-y_0, h) + m_y G_2(x-x_0, y-y_0, h) + m_z G_3(x-x_0, y-y_0, h). \quad (6)$$

The advantage of the first approach is mostly shorter computational times (this is specific for the single-dipole case), with the tradeoff that the larger search space may lead to trapping at local minima and sub-optimal solutions, particularly when the true source distribution cannot be exactly represented by a single dipole. On the other hand, the second approach yields solutions in a smaller number of iterations (although usually taking longer time) and is less prone to trapping at local minima.

In both cases, the optimization problem is solved multiple times using different initial guesses for the optimization parameters at each time that are obtained via random perturbations of the nominal values (with typical perturbations of 5-20% of initial guesses). The final error (residual) associated with each solution is compared and the one with the smallest error is chosen. This procedure helps ensure that local minima are avoided and that the solution that best approximates the experimental data is found. Usually, we compute the optimization problem 20-100 times, obtaining final solutions within 30-60 s on a moderately fast PC with a single Intel Quad Core i7-950 CPU and 12 GB of RAM, depending on size of the field map and how far the nominal initial guess is from the true solution. We typically use hundreds to thousands of data points to recover the three net moment components, resulting in a greatly overdetermined problem that increases the robustness of the solution in the presence of higher noise levels. Clearly, the larger the number of data points, the greater the time to complete each iteration of the optimization procedure. In general, noisy field data require using fine-sampled field maps with a larger number of data points in the optimization so as to achieve adequate accuracy in the moment estimates.

It is often advantageous to run the optimization algorithm a single time while observing the output of the model during the initial iterations, particularly prior to starting to processing maps associated with demagnetization or remagnetization sequences of a sample. This allows us to manually adjust the initial guess for the optimization parameters so that they are not very far off from the true configuration of the experiment, speeding up the overall optimization procedure and helping ensure convergence. For instance, good estimates for the horizontal coordinates $(x_0, y_0)$ of the dipole can be directly obtained from the location with maximum field strength in the total field map computed from the measured normal field component map. Estimates for the distance between the measurement plane and the sources (also called liftoff distance or sensor-to-sample distance), $h$, can be obtained by measuring standard samples such as thin current-carrying wires and small magnetized dots, or even by optical measurements depending on the type of magnetometer used [*Baudenbacher et al.*, 2002; *Hankard et al.*, 2009; *Lima et al.*, 2014].

When solving the six-parameter optimization problem (Case 1), estimates for the moment components can be more easily obtained in spherical coordinates—moment strength, inclination, and declination—and then converted back to rectangular coordinates. Inclination can be roughly estimated by visually comparing the greatest positive and negative values of the measured field $b_z$. Approximately equal values are indicative of zero inclination, whereas predominantly positive or negative values correspond to +90° or -90°, respectively (or else to -90° or +90°, respectively, depending on the convention used). Intermediate inclinations can be reasonably estimated from the ratio between the greatest positive and negative values. Declination can be estimated from the angle between the $y$ axis and the line connecting the greatest positive and negative values of the field map. Lastly, an order-of-magnitude estimate of the strength can be found by comparing the overall magnitude of the field values in the experimental and model field maps.

We emphasize that only coarse estimates of the optimization parameters are required to achieve rapid convergence, and they are usually only necessary for the first step in a demagnetization or remagnetization sequence. In this case, it is also beneficial to use the solution of a particular sequence step as the initial guess for the subsequent step, as this can speed up the overall processing time. Henceforth we focus on the Case 1 algorithm, which was the approach chosen to process all the synthetic and experimental data shown in this paper.

*3.2 - Optimization algorithm*

To solve the optimization problem, we utilized a nonlinear least-squares algorithm with no added bound constraints based on the subspace trust-region method described in [*Coleman and Li*, 1994; 1996], which is implemented in MATLAB® by the function *lsqnonlin*. For the typical field strengths observed in scanning SQUID microscopy and field maps expressed in nano-tesla (nT) units, we used as stopping criteria (i) a tolerance for changes in the value of the objective function of $1\times10^{-12}$, (ii) a tolerance for changes in the size of a step of $1\times10^{-14}$, (iii) a maximum number of iterations of 3000, and (iv) a maximum number of function evaluations of 6000.

We also carried out a number of tests using the Nelder-Mead simplex algorithm [*Lagarias et al.*, 1998], which is implemented in MATLAB by the function *fminsearch*. However, we did not find any appreciable increase in the accuracy of solutions that would warrant the longer convergence times associated with that algorithm.

*3.3 - Uniqueness*

The question of uniqueness of the solution to the net moment inverse problem is intrinsically related to the existence of magnetically silent sources (annihilators) with nonzero moments. In order to uniquely recover the net moment from magnetic field measurements, there cannot exist any magnetization distribution with nonzero moment in a given class of magnetizations that produces no external magnetic field. Otherwise, two magnetizations with different net moments would be indistinguishable as they would produce the same observable field. Usually, a comprehensive characterization of silent sources in a general setting involves sophisticated mathematics [*Baratchart et al.*, 2013; *Parker*, 1994], but this issue is greatly simplified in the case of a single dipolar source. In this situation, it is easy to see from (3) and its generalization for *N* measurement points that there does not exist a nonzero dipole that produces a zero field everywhere on the plane $z=h$. Thus, the solution to problem (4) is inherently unique provided that the field is adequately sampled on the plane, such that the discretization of the problem is not an issue.

## 4. Synthetic samples and sensitivity analysis

As in our approach, standard rock magnetometers infer magnetic moment by assuming the measured samples are dipolar or nearly dipolar. The development of these magnetometers was guided by modeling the magnetic fields of samples of various sizes and shapes, with the sizes and placement of their coils optimized such that the inferred moments were achieved accuracy to a few % for cm-scale uniformly magnetized cylindrical and cubic samples [*Collinson*, 1983]. Following this approach, here we determine the maximum size of uniformly magnetized samples whose moments can be accurately retrieved using our technique given the typical sensor-sample distance of ~100 μm encountered in SQUID microscopy. Specifically, we tested our moment estimation technique with synthetic magnetic field maps obtained using three different types of sources — single magnetic dipole, uniformly magnetized square measuring $50 \times 50$ μm$^2$, and uniformly magnetized cube of $50 \times 50 \times 50$ μm$^3$ — and various combinations of noise level and sensor-to-sample distance. Gaussian white noise was added to field maps to simulate measurements under actual experimental conditions. For the noisy cases, we generated 15 field maps for each source type - noise level combination. This enabled us to determine the statistical dispersion in the recovered parameters due to different realizations of the noise stochastic process.

Regarding the test sources chosen, the magnetic dipole allows us to assess how sensitive the algorithm is to the initial guesses for the model parameters and to noise, because the same source is used in the forward and inverse problems. By using the uniformly magnetized square and cube test sources, we can then determine how performance is degraded as the source strays from a purely dipolar behavior and how noise further impacts the magnetic moment estimates. In this case, the sensor-to-sample distance intrinsically controls the proportion of higher order multipole terms (e.g., quadrupole, octupole, etc.) introduced in the forward model relative to the dipolar term. Whereas all these higher order terms have zero net moment by virtue of the orthogonality properties of spherical harmonics, they may still negatively affect the net moment estimates: given that those estimates are based on the matching of experimental and model field maps, the dipole model parameters will be tuned so as to best match the whole experimental map and not just the dipole component present in it. Therefore, care should be taken to ensure that the dipole term is indeed dominant in the experimental data. If necessary, the magnetic field map should be upward continued to decrease the contribution of higher order terms. The prevalence of the dipole term can usually be evaluated by analyzing the residual map and observing how correlated features eventually present in it change as liftoff distance varies. Notice that this upward continuation approach often yields better results than measuring the sample from a greater distance, as higher signal-to-noise ratio (SNR) can be achieved and contamination of the field map by background sources and adjacent contamination is minimized.

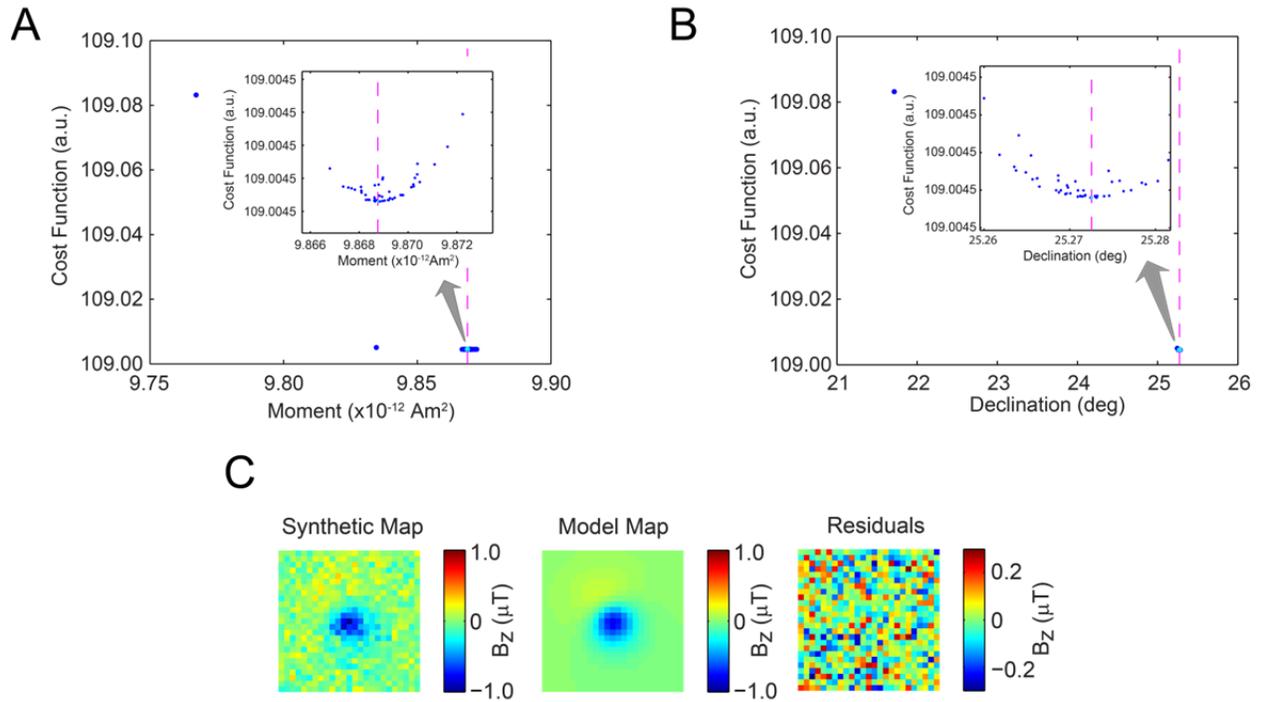

**Figure 2:** Optimization procedure for estimating the net moment of the source. The optimization problem was solved 50 times, each with different initial guesses for the model parameters. The solution with the smallest cost function value (i.e., smallest residual) was then selected as the estimated magnetic moment. **(A)** Plot of the recovered moment magnitude for each of the 50 calculated solutions with their corresponding costs. Each dark blue dot represents a single solution. Inset shows detail of the region containing solutions with the smallest costs. Dashed magenta line and corresponding light blue dot indicate the best solution to the optimization problem. **(B)** Plot of the recovered declination for each of the 50 calculated solutions with their corresponding costs. Other model parameters exhibit analogous behavior. **(C)** Left: synthetic field map of a uniformly magnetized $50 \times 50 \times 50$ µm$^3$ cube measured on a planar grid 100 µm above it. White noise was added to yield a very poor SNR of 0 dB (i.e., a 1:1 proportion of noise and signal). Center: field map of the magnetic dipole that best fits the data, which was found using the abovementioned optimization procedure. Right: difference between synthetic and model maps, showing that they differ essentially by the noise component.

For a given field map, the optimization was solved 50 times, each one with initial parameter estimates that consisted of random perturbations of the initial guess by as much as 10% (up to ±4° and ±8° for the inclination and declination parameters, respectively). As explained in Section 3, the net moment estimate was obtained from the optimization solution with the smallest residual (Fig. 2). For each combination of source type and SNR, we computed a total of 15 net moment estimates: in the noiseless case, we used identical field maps, whereas in the noisy cases, we used maps with different realizations of the random noise. We then computed the mean net moment by (vector) averaging the 15 net moment estimates. We also calculated the sample standard deviation and sample mean (or median depending on the case) for the moment magnitude, recovered height, and angular error.

Notice that the noiseless maps allow us to demonstrate the consistency of the net moment estimates, given that no scatter should be observed in such estimates when repeatedly inverting identical data if the global minimum of the objective function is being effectively reached during the optimization procedure.

We begin by analyzing the noiseless case for all three synthetic sources (Figs. 3A, 4A and 5A). As expected, the scatter in the recovered quantities — denoted by error bars representing plus or minus one sample standard deviation — is negligible when no noise is present. Notice that the standard error of the mean (SEM), which measures the standard deviation of the error in the sample mean relative to the true mean, can be obtained by scaling the sample standard deviation by $1/\sqrt{N}$, where $N$ is the sample size. In our case, this corresponds to shrinking the error bars by a factor of ≈3.9 to represent the SEM. Owing to the skewed nature of the distribution of angular error in the net moment estimates, we display the median and first and third quartiles in that case.

There is no dependence of estimated moment magnitude and direction on liftoff distance for the dipole, as expected (Figs. 3A and 4A). There is also no deviation in recovered liftoff with respect to the true liftoff distance (Fig. 5A). However, there is noticeable dependence of those quantities for the uniformly magnetized square and cube sources. Such dependence is stronger for the square, which should be expected given that the center of the cube lies deeper than that of the square (by definition, the liftoff distance is measured between the magnetic sensor and the top of the source distribution/sample). In particular, the recovered dipole lies deeper than the actual test source to compensate for the slower spatial decay of the magnetic field in those cases.

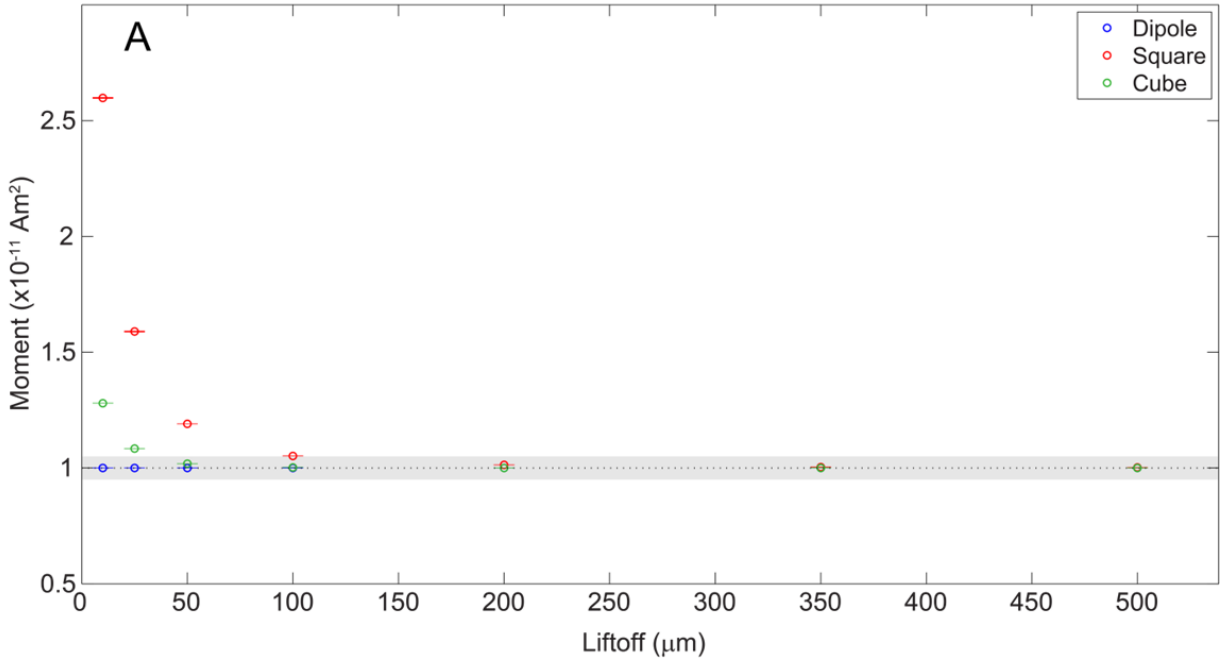

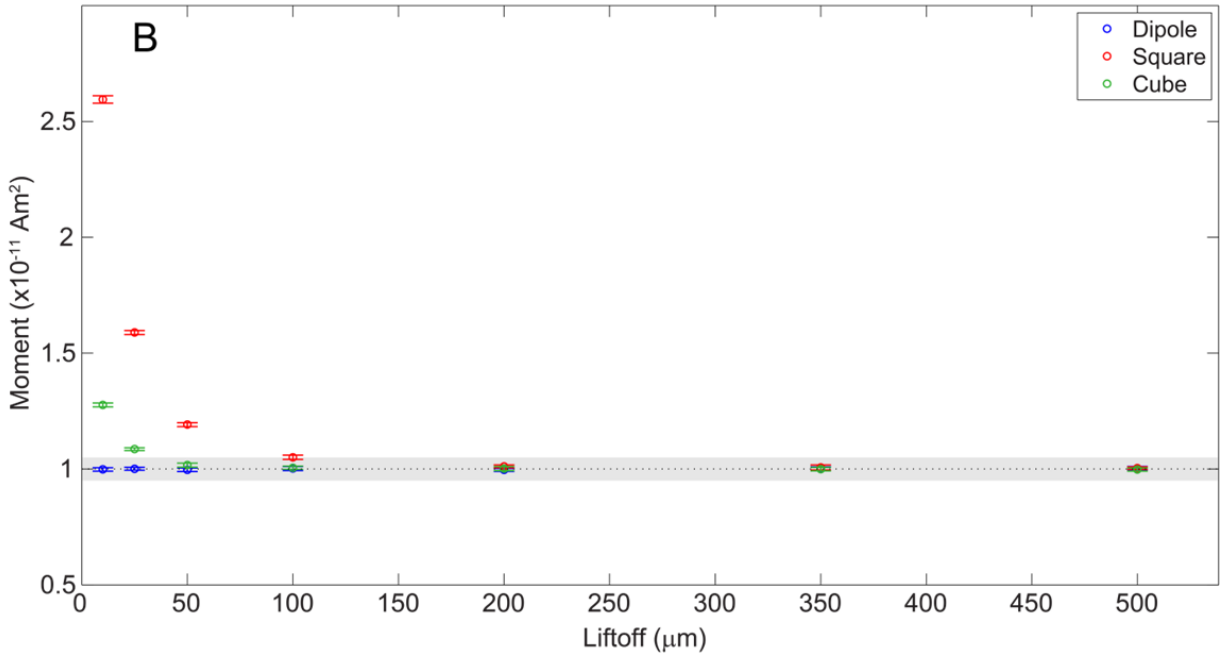

**Figure 3:** Magnitude of estimated net moments as a function of liftoff distance for all three test sources and different noise levels. Each colored circle represents the magnitude of the mean net moment (the mean of the magnitudes of estimated moments was omitted for clarity purposes, as it virtually coincides with the magnitude of the mean net moment). The error bars represent plus or minus one sample standard deviation. **(A)** Noiseless case (SNR = ∞). **(B)** 10:1 proportion of signal and noise (SNR = 20 dB).

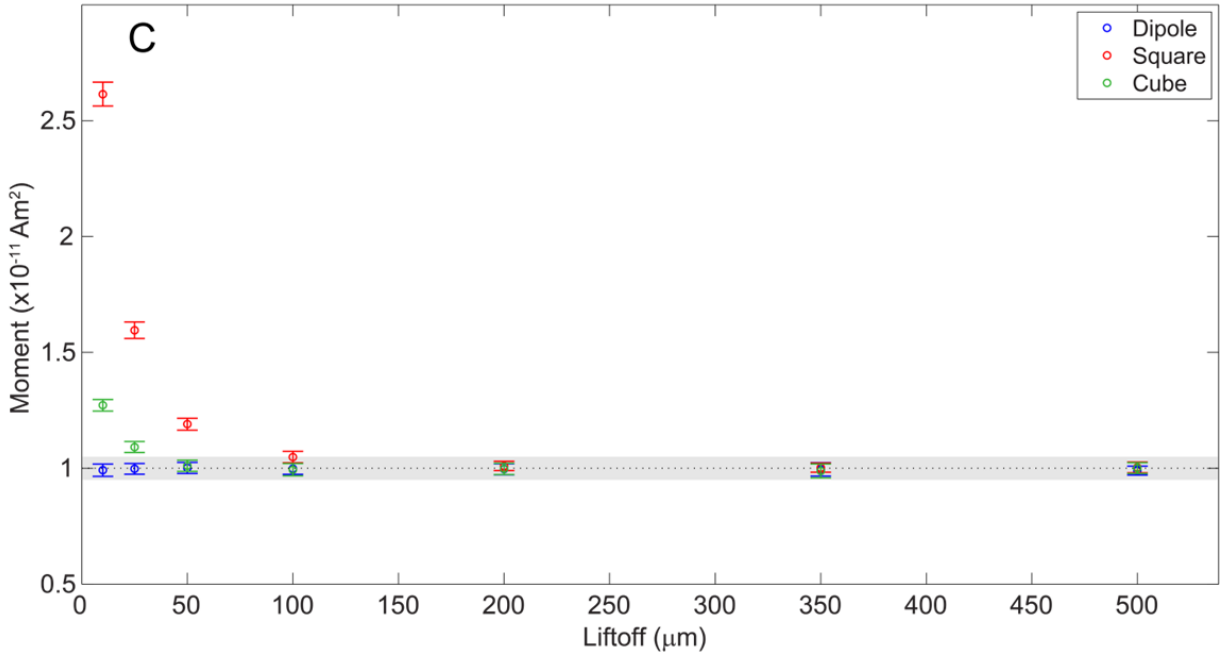

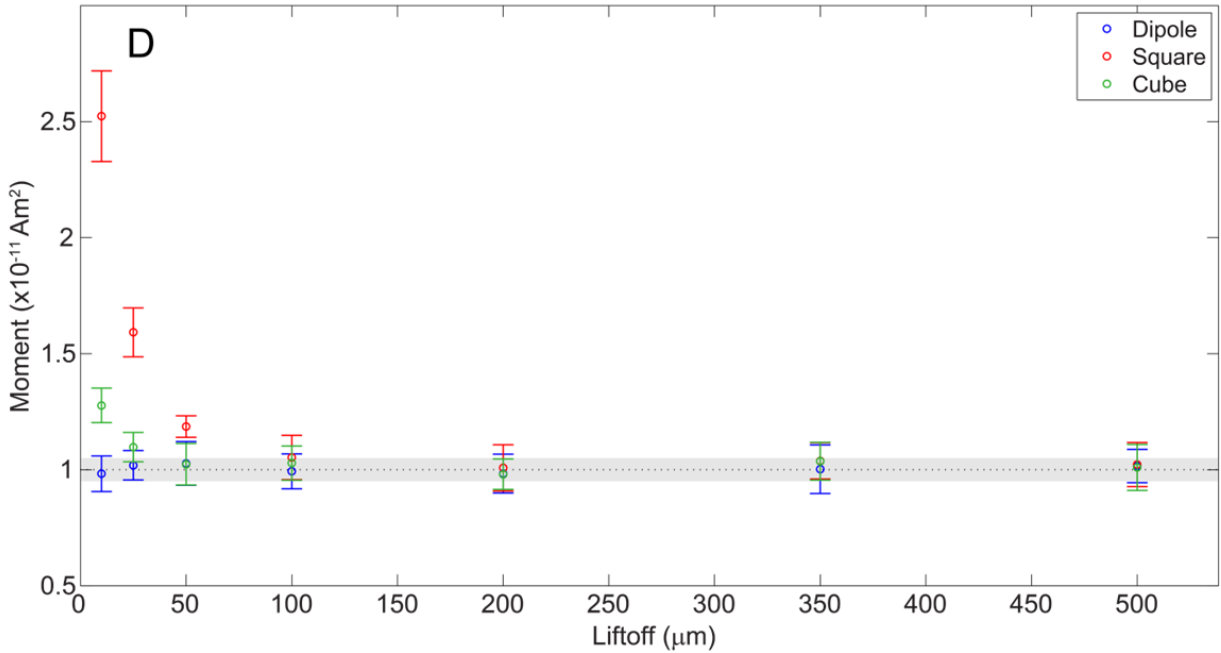

**Figure 3 (cont): (C)** 3:1 proportion of signal and noise (SNR = 10 dB). **(D)** 1:1 proportion of signal and noise (SNR = 0 dB). Black dashed line indicates the true magnitude of the moment, and the gray rectangle denotes the region in the plot where deviations from the true magnitude are smaller than or equal to 5%.

As a consequence, the magnitude of the moment has to increase so as to match the strength of the magnetic field. The resulting error in the estimated moment, especially in the magnitude, can be quite large (>250 %) when the liftoff distance is smaller than the source dimensions (i.e., < 50 um for square plate). For larger liftoffs (> 100 um), the dipole term becomes dominant and accuracy improves rapidly.

Introducing noise in the field maps results in scatter in the net moment estimates and in other model parameters. At moderate noise levels (20 dB SNR or 10:1 proportion between signal and noise), a small scatter of a few percent is noticeable in the moment magnitude (Fig. 3B) and of about a degree in the direction (Fig. 4B). The scatter in the recovered liftoff is comparatively smaller and barely noticeable (Fig. 5B). Increasing the noise level to yield an SNR of 10 dB (i.e., 3.2:1 proportion of signal and noise) has the effect of increasing the scatter by a comparable amount (Figs. 3C, 4C, and 5C). Finally, for a very poor SNR of 0 dB (i.e., equal proportion of signal and noise) the scatter increases further by another factor of $\approx 3$. Notice that in all noisy cases, the overall trend of the recovered quantities observed in the noiseless case for the three sources is preserved.

An important point that can be seen in Figs. 3-5 is that the mean net moment typically yields better estimates of the true moment's magnitude and direction than directly computing the mean or median of the individual parameters. For example, the angular error of the mean net moment is noticeably smaller than the median of the angular errors of the individual solutions. Whereas in paleomagnetic studies it may not be practical to repeatedly map a sample 15 times (for each magnetization/demagnetization step), as was done in this computational experiment, it is nevertheless beneficial to make a few repeated maps so as to improve the accuracy of the net moment estimates through vector averaging, particularly when measuring very weak samples with degraded signal-to-noise ratios.

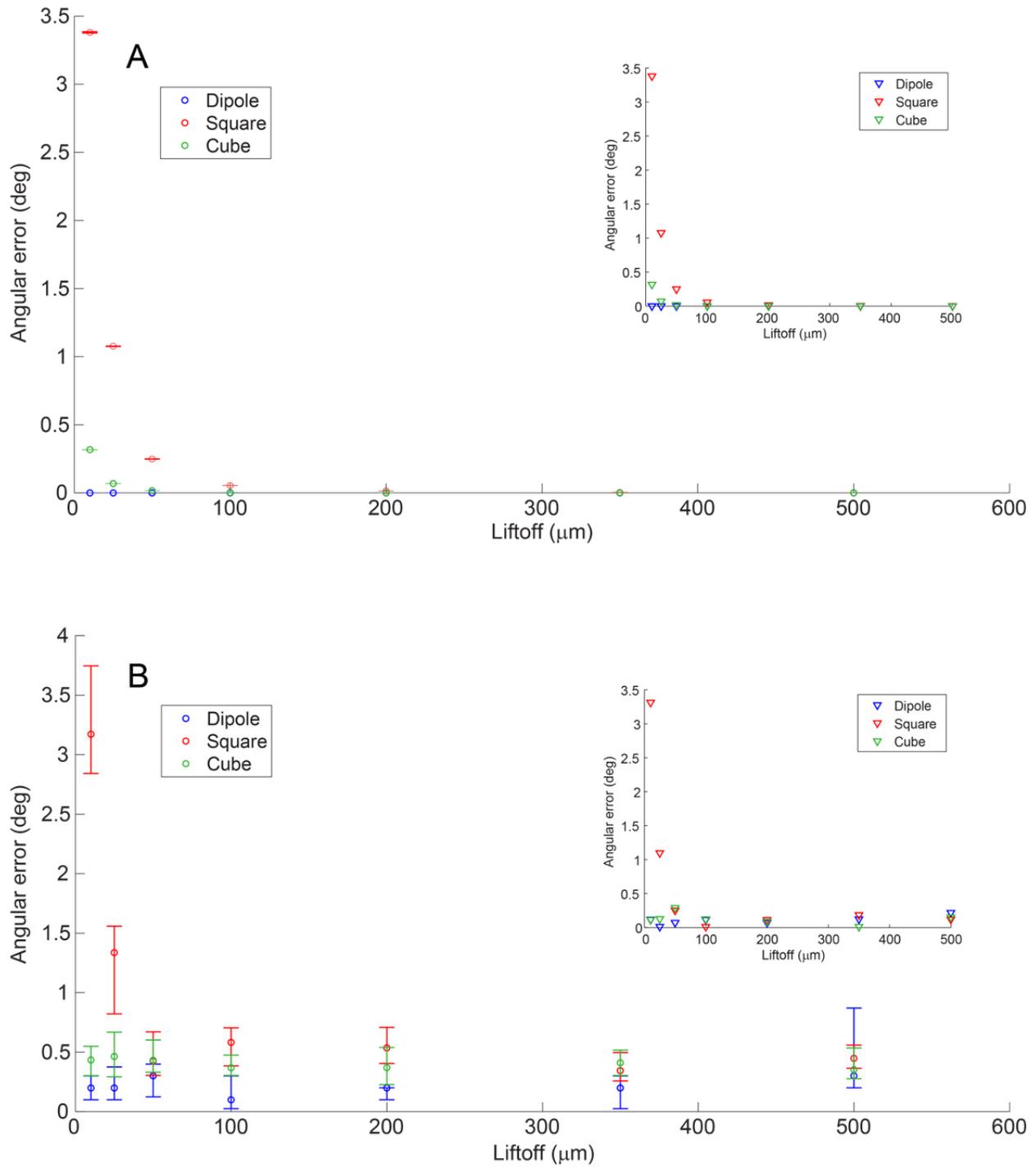

**Figure 4:** Error in the direction of the estimated net moments as a function of liftoff distance for all three test sources and different noise levels. Each colored circle represents the sample median of the angular error. The error bars denote the first and third sample quartiles. Colored triangles in the insets show the angular error of the mean net moment, which is typically smaller than the sample median. **(A)** Noiseless case (SNR = ∞). **(B)** 10:1 proportion of signal and noise (SNR = 20 dB).

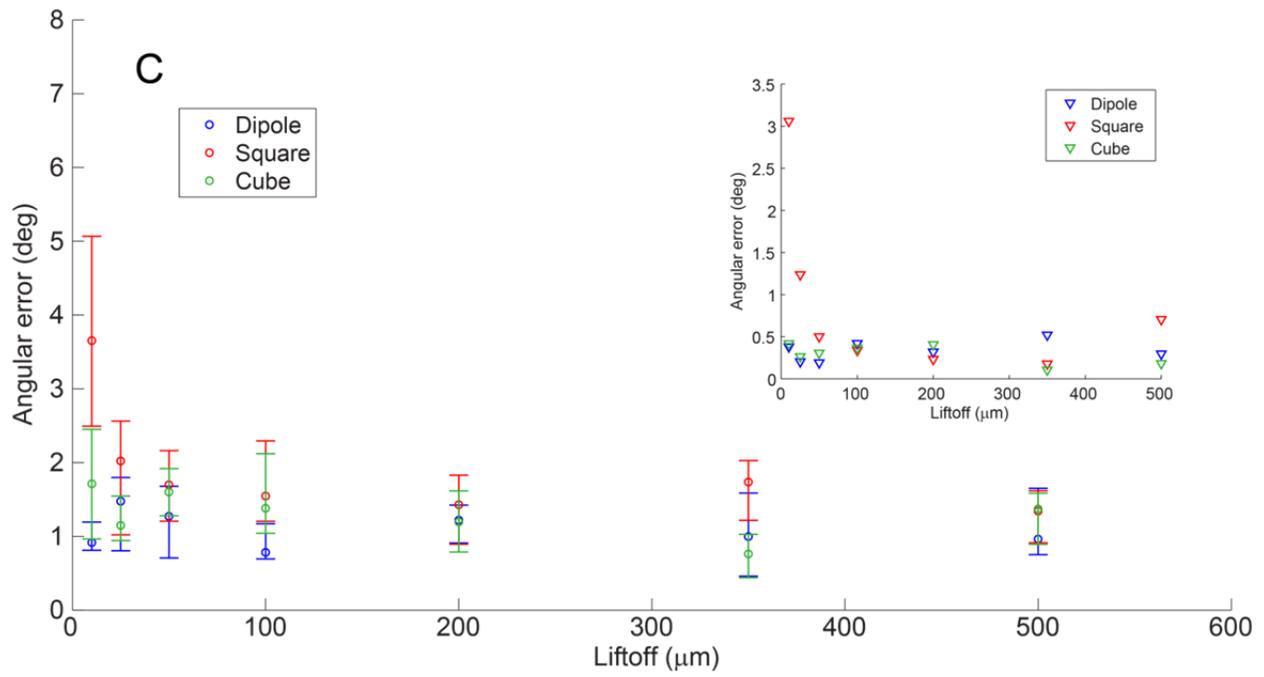

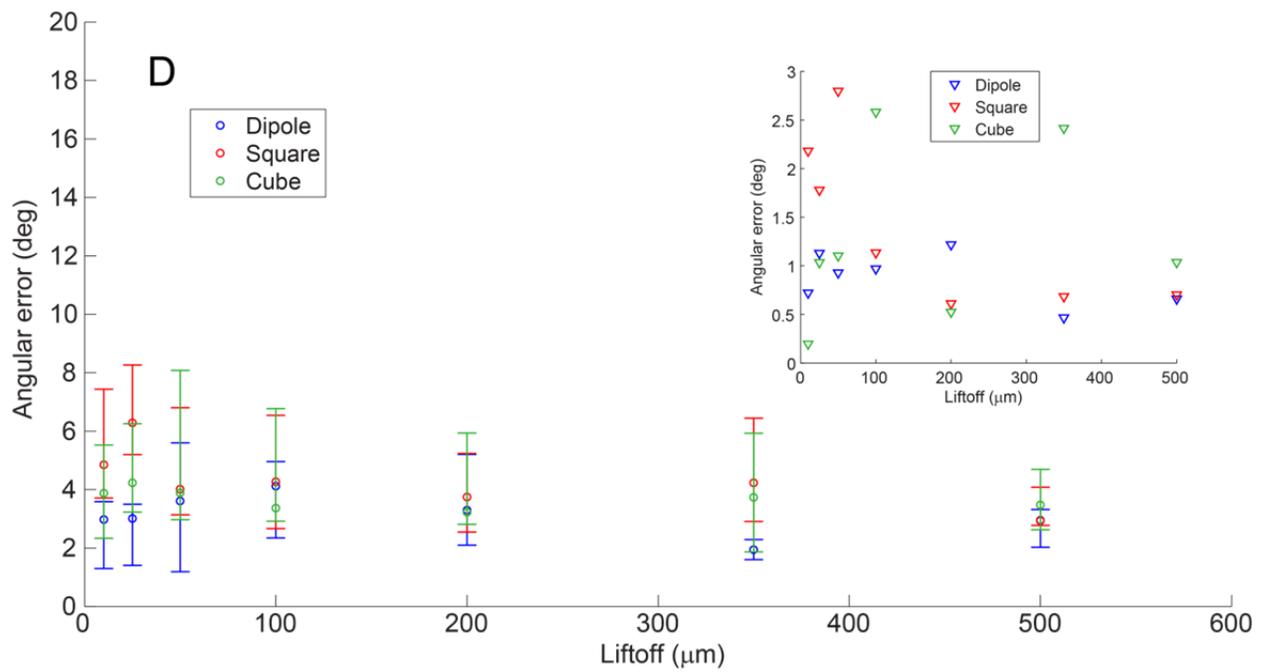

**Figure 4 (cont): (C)** 3:1 proportion of signal and noise (SNR = 10 dB). **(D)** 1:1 proportion of signal and noise (SNR = 0 dB).

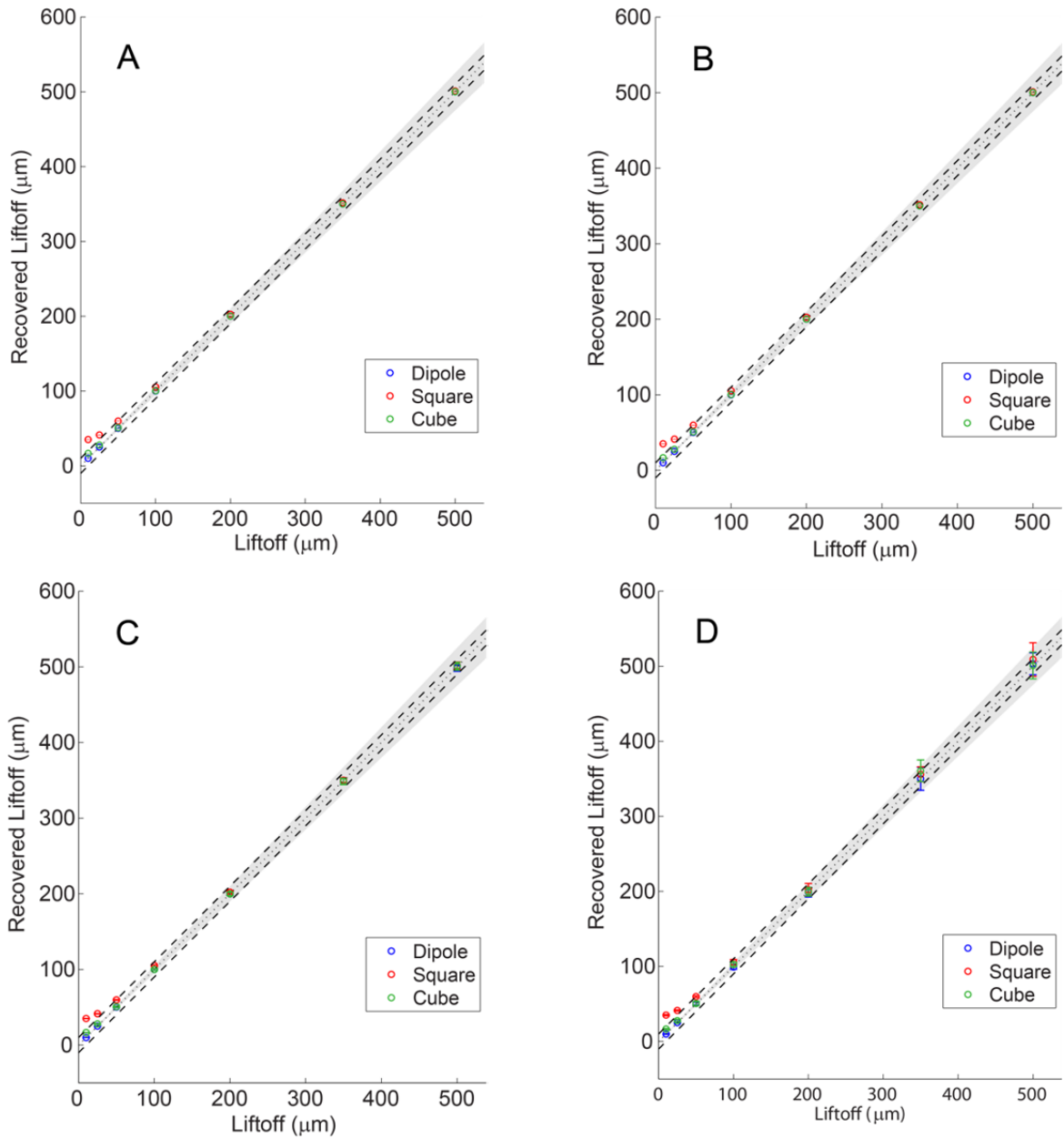

**Figure 5:** Recovered liftoff as a function of liftoff distance for all three test sources and different noise levels. Each colored circle represents the recovered liftoff. The error bars represent plus or minus one sample standard deviation. **(A)** Noiseless case (SNR = ∞). **(B)** 10:1 proportion of signal and noise (SNR = 20 dB). **(C)** 3:1 proportion of signal and noise (SNR = 10 dB). **(D)** 1:1 proportion of signal and noise (SNR = 0 dB). Black dotted line indicates the true liftoff, and the gray area denotes the region in the plot where deviations from the true liftoff are smaller than or equal to 5%. Two black dashed lines represent ±10 μm deviation from the true liftoff.

## 5. Application to geological samples

To experimentally validate our technique, we imparted controlled magnetizations on actual geological samples with dipolar magnetization characteristic. We measured those samples using both a commercial SQUID rock magnetometer (2G Enterprises 755 SRM) (sensitivity $1\times10^{-12}$ Am$^2$) and the SQUID microscope (sensitivity $1\times10^{-15}$ Am$^2$) housed in the MIT Paleomagnetism Laboratory [*Fong et al.*, 2005]. (In this section, no vector averaging of moment estimates was performed, so as to provide a fair and direct comparison between our technique and standard SQUID rock magnetometry.)

*5.1 – Impact spherule from Lonar crater, India*

We began by magnetizing a ~300 μm impact melt spherule from Lonar crater in India [*Weiss et al.*, 2010] by imparting a 200 mT isothermal remanent magnetization (IRM). The sample was then mounted on an acid-washed 2.5 cm diameter quartz disc using cyanoacrylate (superglue), which are both magnetically clean. This yielded a moment of $5.8\times10^{-9}$ Am$^2$, >1000× the detection limit of the 2G SRM. This sample was progressively demagnetized using alternating field (AF) demagnetization methods. For each demagnetization step, the sample was measured on both instruments and the results compared (Figs. 6A and 6B).

The sample's moment decreased down to $4.0\times10^{-11}$ Am$^2$, always remaining above the sensitivity limits of both magnetometers. This high moment explains why the measurements of the two magnetometers agree well for most of the demagnetization sequence. However, there are noticeable discrepancies in direction for demagnetization steps above 100 mT even though the sample moment at these steps is still $3.8\times10^{-10}$ – $4.0\times10^{-11}$ Am$^2$. In this AF range, the directional data from the SQUID microscope measurements show a ~60° change in declination with almost no change in inclination between the first (i.e., IRM) and last (AF 200 mT) steps. In contrast, the 2G data imply a hemisphere change and inclination shallowing in the last demagnetization steps.

To determine which instrument is yielding accurate data, we look at the field maps of the first (Fig. 6C) and last (Fig. 6D) steps in the sequence. It is very clear that there is no visually perceptible change in inclination whereas declination varies by approximately 60°. (Information about declination can be obtained from the line connecting the maximum and minimum field in a map; inclination is associated with the ratio between the maximum and minimum fields in a map, with zero inclination corresponding to a ratio of 1.) This suggests that, in this measurement scenario, the SQUID microscope data are much more accurate than the 2G data.

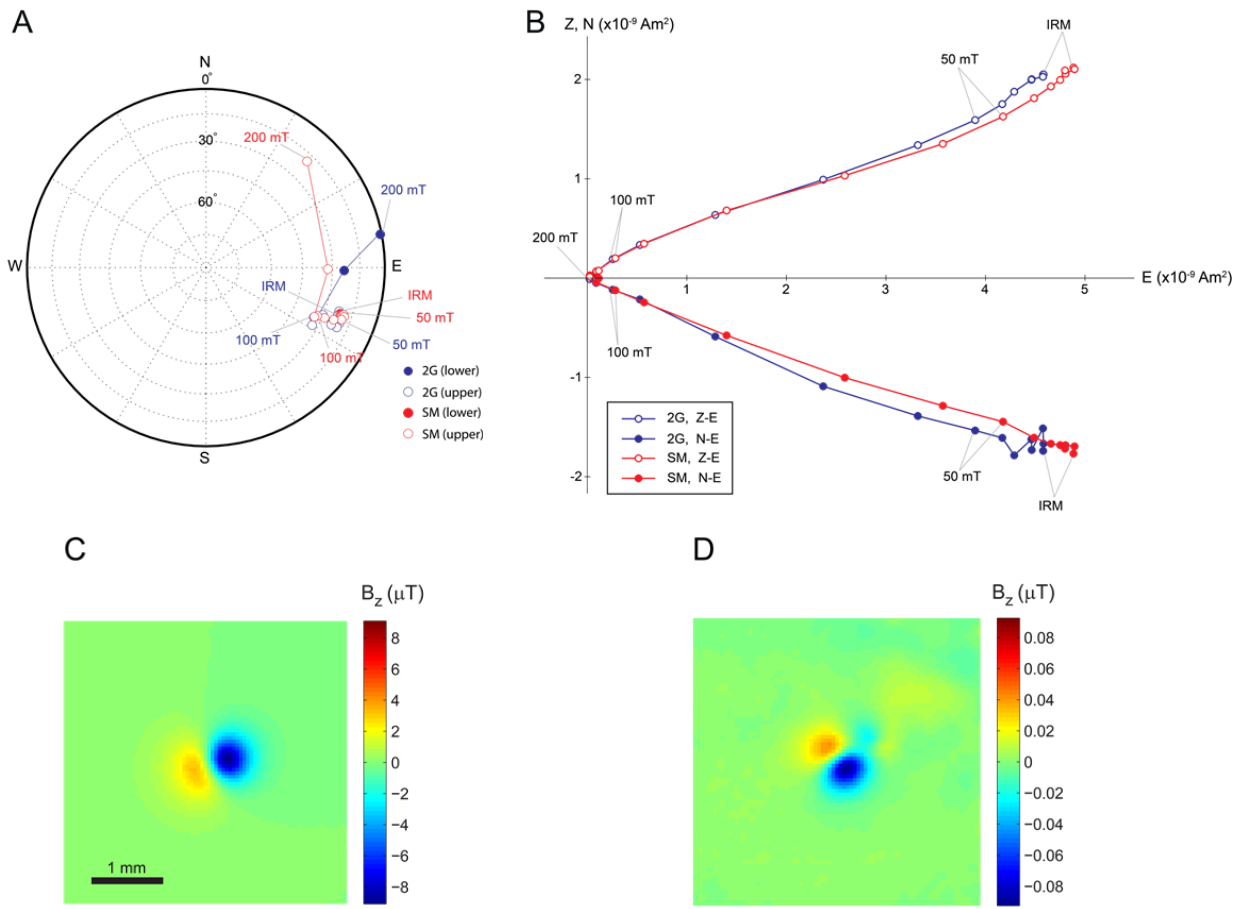

**Figure 6:** AF demagnetization of an impact melt spherule from Lonar crater in India. We imparted a 200 mT IRM to the spherule prior to mounting it on a quartz disc and subsequently carrying out the AF demagnetization steps. At each step, the spherule was measured on the 2G SRM and on our SQUID microscope. **(A)** Net moment directional data shown on an equal area stereoplot (SQUID microscope data shown in red, 2G data shown in blue). **(B)** Up-east and north-east projections of the endpoints of the net moment vector shown (SQUID microscope data shown in red, 2G data shown in blue). **(C)** Vertical component (i.e., normal to the sample mount) of the sample's magnetic field after an IRM 0.2 T was imparted. **(D)** Vertical component of the sample's field after the final AF demagnetization step (200 mT peak field). Notice the ~60° change in declination between the two steps with almost no change in inclination. (Magnetic field maps were measured ~200 μm above the sample.)

The reason for the superior performance of the SQUID microscope can be observed from Fig 6D: areas with weaker magnetization surround the spherule, likely from impurities in the quartz disc and dust particles that became trapped in the glue. The mapping area shown is 4 × 4 mm$^2$ and only reveals a small fraction of the contaminating magnetization that could be present in a 2.5 cm diameter quartz disc. The small discrepancy in Fig. 6B in the beginning of the demagnetization sequence is likely attributed to contaminating magnetization. Such secondary sources have very little effect on the net moment estimates obtained from the SQUID microscopy data but directly affect the 2G measurements. This is an enormous advantage of magnetic microscopy over standard rock magnetometry when dealing with very weak samples: contamination in the sample holder is visually evident and its effects can be minimized in the vast majority of cases. On the other hand, it can be extremely difficult to ascertain whether standard SQUID rock magnetometry data are partially biased or even dominated by secondary sources when measuring very weak samples.

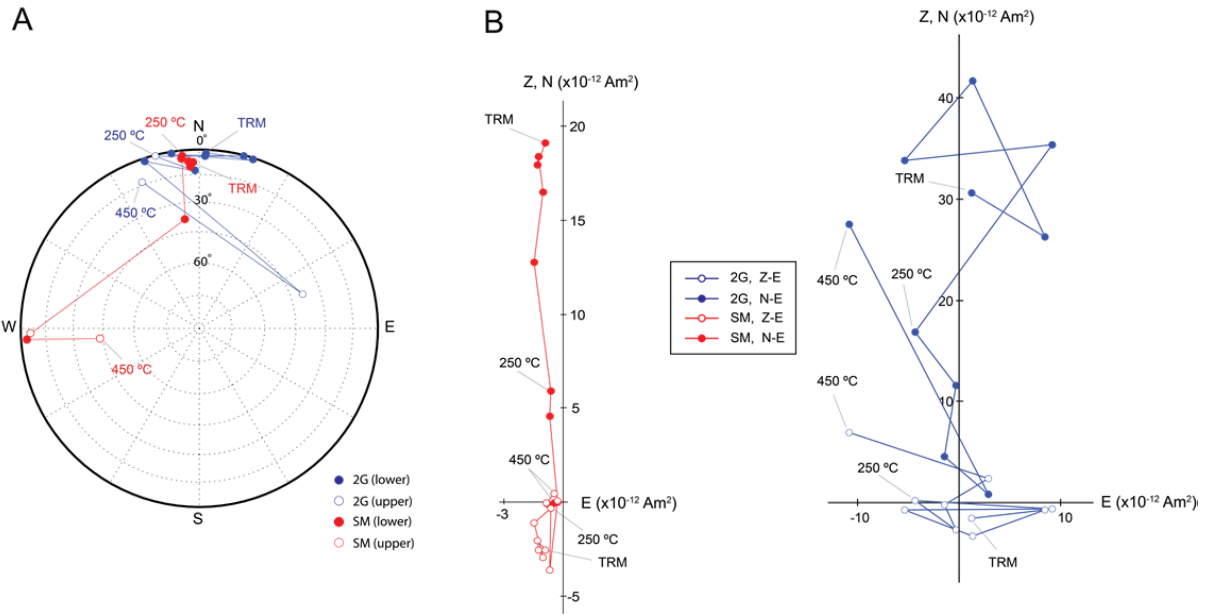

**Figure 7:** Thermal demagnetization of a small clump of cutting dust from the Millbillillie eucrite. We imparted a 50 µT TRM at 580 °C to the sample prior to carrying out a sequence of thermal demagnetization steps. At each step, the sample was measured on the 2G rock magnetometer and on our SQUID microscope magnetometer. **(A)** Net moment directional data shown on an equal area stereoplot (SQUID microscope data shown in red, 2G data shown in blue). **(B)** Up-east and north-east projections of the endpoints of the net moment vector shown (SQUID microscope data shown in red, 2G data shown in blue). (Magnetic field was mapped ~200 µm above the sample.)

*5.2 – Millbillillie eucrite*

We then proceeded to measure a weaker sample whose moment dropped below the noise limit of the 2G SRM during a demagnetization sequence (from $1.9 \times 10^{-11}$ Am$^2$ down to $6.6 \times 10^{-13}$ Am$^2$). Cutting dust from the Millbillillie eucrite [*Fu et al.*, 2012a] was collected during sawing of the meteorite and packed into a small clump on a 2.5 cm diameter quartz disc using silver paste designed for scanning electron microscopy applications as a non-magnetic adhesive. A TRM was imparted by heating the sample to 580 °C while applying a 50 µT dc magnetic field, followed by thermal demagnetization steps up to 450 °C (Fig. 7). The 2G SRM data are noisy and do not trend to the origin, which we attribute to a combination of contamination sources and instrument noise. In contrast, the SQUID microscope data are much cleaner and clearly trend to the origin up to temperatures of 450 °C. Because the present experiment only involved thermal demagnetization, it demonstrates that the observed discrepancies and scatter in the data cannot be attributed to spurious anhysteretic remanent magnetization (ARM) noise during AF demagnetization.

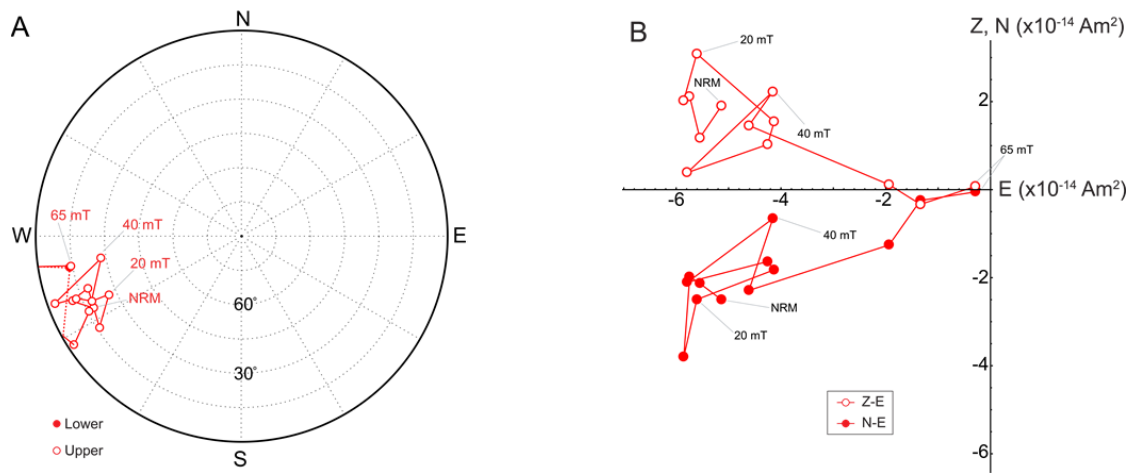

**Figure 8:** AF demagnetization sequence of a Jack Hills zircon measured with our SQUID microscope. **(A)** Net moment directional data shown on an equal area stereoplot. **(B)** Up-east and north-east projections of the endpoints of the net moment vector. (Magnetic field was mapped ~200 µm above the sample.)

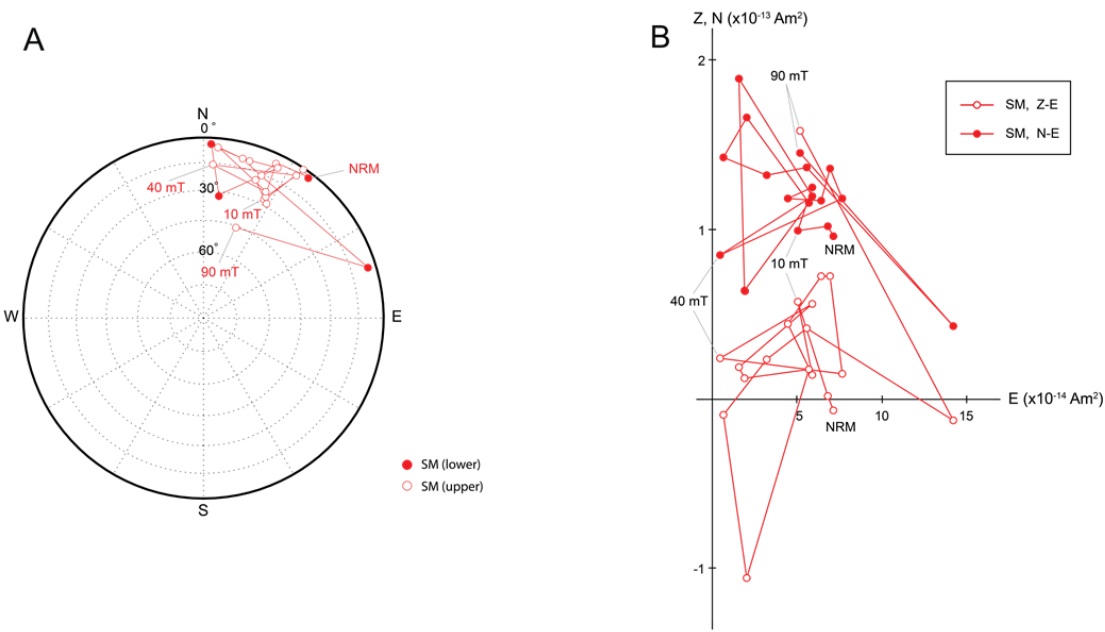

**Figure 9:** AF demagnetization sequence of a Jack Hills zircon measured with our SQUID microscope. **(A)** Net moment directional data shown on an equal area stereoplot. **(B)** Up-east and north-east projections of the endpoints of the net moment vector shown. (Magnetic field was mapped ~200 μm above the sample.)

*5.3 – Detrital zircons from the Jack Hills, Western Australia*

Having established that the technique agrees with independent measurements for stronger samples and that it yields robust data for weaker ones, we proceeded to apply the technique to ultra-weak samples that cannot be detected using standard rock magnetometers. Detrital zircon crystals from the Jack Hills in Western Australia are an important target as they may preserve records of the origin and earliest evolution of the geodynamo (Weiss et al. 2015). Their magnetizations typically fall below the detection limit of commercial rock magnetometers and exhibit dipolar characteristics at the spatial scale of SQUID microscopy.

To demonstrate the performance of our ultra-high sensitivity moment magnetometry technique, we conducted AF demagnetization data on two detrital zircons from the Jack Hills (Figs. 8 and 9) extracted from the Hadean zircon sampling site at Erawandoo Hill (site EHJH5 of [*Weiss et al.*, 2015]) . Given their very weak moments, each zircon was carefully mounted on a separate 2.5 cm diameter quartz disc using cyanoacrylate. This procedure was performed in a clean room using non-magnetic ceramic tools to minimize any chance of contamination by dust and other spurious sources of magnetic field. The first zircon clearly shows an origin-trending pattern with good directional stability (Fig. 8), starting with an NRM moment of $5.9\times10^{-14}$ Am$^2$ that demagnetized to $4.0\times10^{-15}$ Am$^2$ by the AF 65 mT step. Moderate scatter in the data is predominantly due to the zircon's behavior under AF demagnetization and cannot be attributed to instrument noise. This point is confirmed by the demagnetization of a second zircon, which exhibits larger scatter and no decay despite having slightly stronger moment magnitude ($\sim1\times10^{-13}$ Am$^2$ — cf. Fig. 9). It is clear that the observed differences in demagnetization pattern indeed stem from the samples and are likely due to the very small quantities of ferromagnetic material present in them.

*5.4 – Other ultra-weak magnetic sources*

To confirm that we are still above the detection limit of the technique we measured an ultra-weak secondary source present in a blob of silver paste on a quartz disc and estimated its net moment, which is just $3.6\times10^{-15}$ Am$^2$ (Fig 10). Despite being extremely faint, analysis of the residuals map reveals that we are still above the instrument's noise floor by at least a factor of 5. This suggests that our actual detection limit — without additional signal processing to filter out instrument noise from the field maps — is in the mid to upper $10^{-16}$ Am$^2$ range. This encompasses the full range of sample moments expected to yield accurate paleodirectional and paleointensity constraints on the ancient field (Section 2).

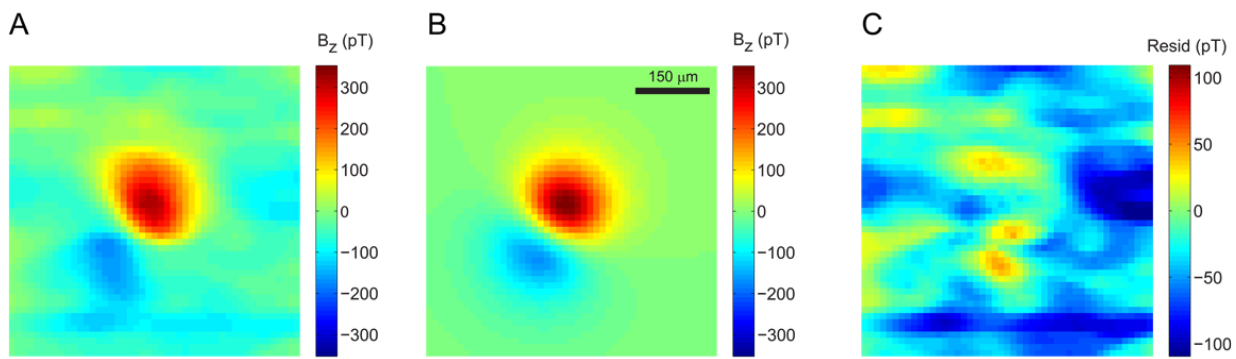

**Figure 10:** Detection and magnetic moment estimation of an ultra-weak magnetic moment consisting of a contaminating pointwise dipolar source present in a blob of silver paste. **(A)** Vertical component (i.e., normal to the sample mount) of the sample's magnetic field. **(B)** Field map of the magnetic dipole that best fits the data. **(C)** Difference between experimental and model maps, showing that they differ by very faint background sources that are still above the instrument noise. The magnetic moment of this source is $3.6\times10^{-15}$ Am$^2$. (Magnetic field was mapped ~200 μm above the sample.)

## 6. Conclusions

We demonstrated that natural geologic samples should contain paleomagnetically useful information down to net moment of 100-1000 times below that of standard rock magnetometers. To measure such samples, we developed an ultra-high sensitivity moment magnetometry technique based on magnetic microscopy that allows us to measure extremely faint magnetic sources with moments as weak as $1\times10^{-15}$ Am$^2$, thereby encompassing the full range of samples expected to be useful for paleomagnetism. We validated the technique by choosing suitable samples that could be independently measured with a commercial superconducting rock magnetometer. This comparison also demonstrated some of the strengths of our technique, which include superior moment sensitivity and a powerful ability to detect contaminating sources while minimizing their effect on the recovered moment. Our analysis with synthetic data revealed that the technique is accurate for dipolar sources even when measurements are contaminated with high levels of noise. It also showed that the major source of error in the net moment estimates is the possible deviation of the sample's magnetization from a magnetic dipole. This can be ameliorated by upward continuing the magnetic field data so as to reduce the contribution of higher order multipole terms and enhance the contribution of the dipole term. Alternatively, more sophisticated source models could potentially be used to improve accuracy (e.g., multiple dipoles, incorporation of quadrupole terms into the modeling, uniformly magnetized areas/volumes) but at the expense of slowing down the algorithm. Although here we applied the technique to data only from SQUID microscopy, our technique can be directly applied to field maps obtained with other scanning magnetic microscopes and magnetic imaging techniques (e.g., magneto-optical imaging [*Uehara et al.*, 2010] and quantum diamond magnetometry [*Hong et al.*, 2013]).

## 7. Acknowledgements and Data

The authors would like to thank the National Science Foundation grant DMS-1521765 for partial support and Mr. Thomas F. Peterson, Jr. for his generous gift to the MIT Paleomagnetism Laboratory that also supported in part this research. All data sets and MATLAB® scripts used in this work are available from the authors upon request.